\documentclass[aps,prl,twocolumn,groupedaddress]{revtex4-1}

\usepackage{graphicx}
\usepackage{amsfonts}
\usepackage{amsmath}
\usepackage{amssymb}
\usepackage{natbib}
\usepackage{graphicx}
\usepackage{color}
\newcommand{\Mg}{$^{24}\text{Mg}^{+}$}
\newcommand{\be}{\begin{equation}}
\newcommand{\ee}{\end{equation}}
\newcommand{\bem}{\begin{multline}}
\newcommand{\ovl}[1]{$\overline{\text{#1}}$}

\begin{document}


\title{Spectroscopy and Directed Transport of Topological Solitons in Crystals of Trapped Ions}



\author{J. Brox}
\affiliation{Albert-Ludwigs-Universit\"at Freiburg, Physikalisches Institut, Hermann-Herder-Strasse 3, 79104 Freiburg, Germany}
\author{P. Kiefer}
\affiliation{Albert-Ludwigs-Universit\"at Freiburg, Physikalisches Institut, Hermann-Herder-Strasse 3, 79104 Freiburg, Germany}
\author{M. Bujak}
\affiliation{Albert-Ludwigs-Universit\"at Freiburg, Physikalisches Institut, Hermann-Herder-Strasse 3, 79104 Freiburg, Germany}
\author{H. Landa}
\email{haggaila@gmail.com}
\affiliation{LPTMS, CNRS, Univ.~Paris-Sud, Universit{\'e} Paris-Saclay, 91405 Orsay, France}
\author{T. Schaetz}
\email{tobias.schaetz@physik.uni-freiburg.de}
\affiliation{Albert-Ludwigs-Universit\"at Freiburg, Physikalisches Institut, Hermann-Herder-Strasse 3, 79104 Freiburg, Germany}

\date{\today}


\begin{abstract}
We study experimentally and theoretically discrete solitons in crystalline structures consisting of several tens of laser-cooled ions confined in a radiofrequency trap.
Resonantly exciting localized, spectrally
gapped vibrational modes of the soliton, a nonlinear mechanism leads to a nonequilibrium
steady state of the continuously cooled crystal. 
We find that the propagation and the escape of the soliton out of its a quasi-one-dimensional channel can be described as a thermal activation mechanism. 
We control the effective temperature of the soliton's collective coordinate by the amplitude of the external excitation. 
Furthermore, the global trapping potential permits controlling the soliton dynamics and realizing directed transport depending on its topological charge.
\end{abstract}


\maketitle

Transport is one of the most basic phenomena studied in physics.
In particular, molecular scale directed transport of matter and energy is of considerable interest \cite{julicher1997modeling,hanggi2009artificial, astumian2016physics}. 
Biological `molecular motors' are submicron machines that consume nondirectional energy to enable directed transport, typically restricted to a one-dimensional (1D) `track'. 
At these scales, the fundamental question arises how the self-propelled motion permits directed transport while competing against stochastic forces.
Similarly, membrane channels, nanopores and nanotubes can be modeled as 1D or quasi-1D systems with a cross section comparable to the size of the transported ion or molecule. 
Here, the entire channel forms the machine controlling the rate and direction of matter transport between two regions against a gradient (e.g. electrochemical), acting in addition to the noise.


\begin{figure}[tb]
	\centering
		\includegraphics[width=3.25in]{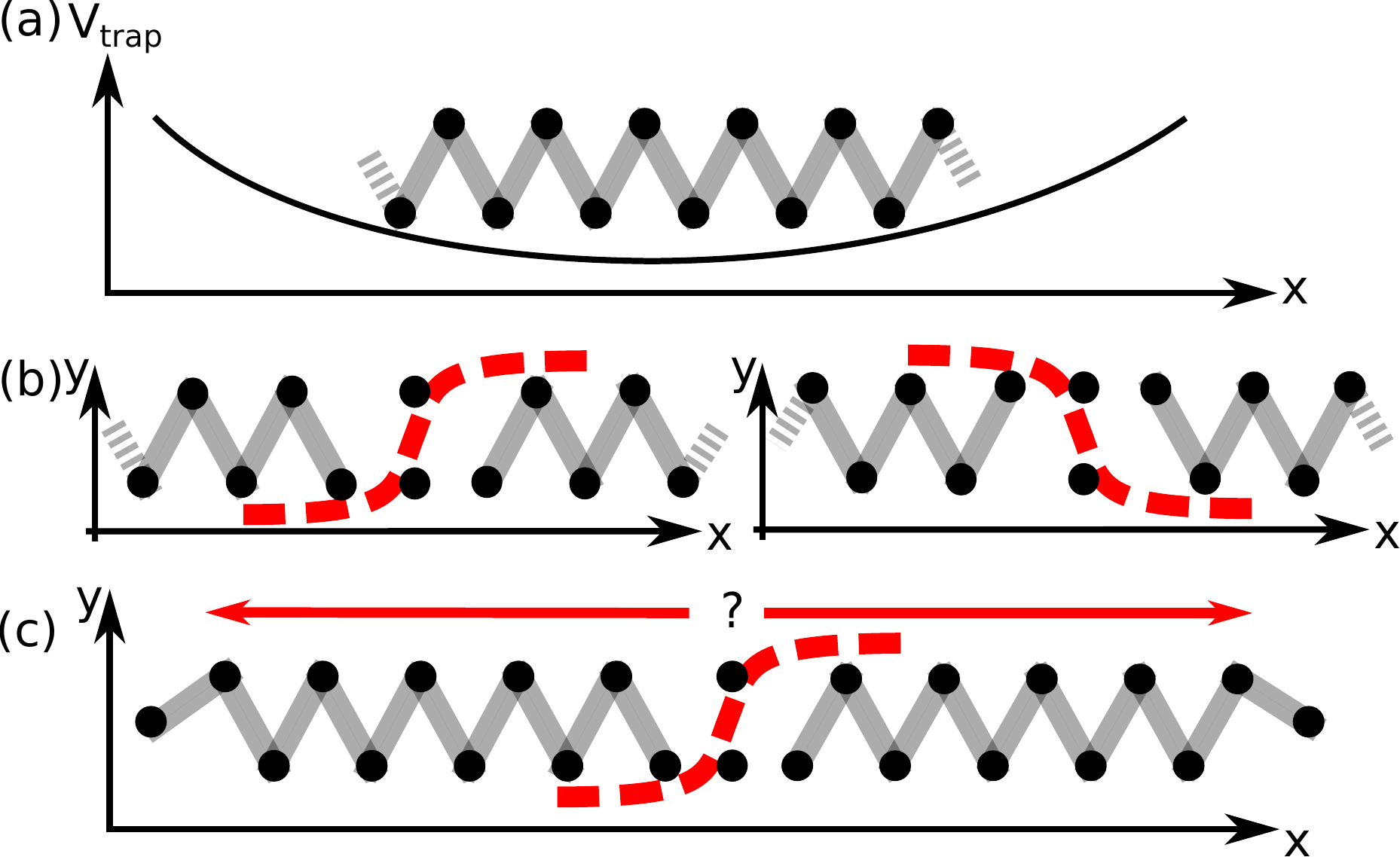}
	\caption{Schematic depiction of discrete solitons and their propagation in a trapped Coulomb crystal. 
	\textbf{(a)} The self-assembled crystal features reflection symmetry and its energy minimum comes in two degenerate configurations, the `zigzag' and its mirror image `\ovl{zigzag}' (only one shown). 
	\textbf{(b)} Realizing both configurations in one crystal requires an interface, a domain wall called `kink' (left) or `\ovl{kink}' (right), which is a discrete soliton, carrying a topological charge of $\pm 1$, illustrated with a dashed red line.
	\textbf{(c)} In this paper we ask whether a resonant global excitation of radial vibrational kink modes can be rectified by the soliton and exploited to conditionally propagate it to the right or left in a noisy environment.
	}
	\label{fig:kinks}
\end{figure}	

A prominent model for such dynamics is the Brownian ratchet, or Brownian motor \cite{denisov2014tunable}.
The basic assumption for the ratchet effect is that all mean forces in the system vanish, and for the Brownian ratchet - the presence of significant noise.
To allow the emergence of nonvanishing (mean) currents, the breaking of a symmetry is required - either spatial, temporal, stochastic or spontaneous.

A natural generalization of the single-particle ratchet to a many-body, nonlinear setting can be achieved with solitons, non-perturbative solutions that manifest a collective particle-like behavior \cite{Rajaraman,dauxois2006physics} (fig.~\ref{fig:kinks}). 
However, solitons are not point particles and have some extension in space, in addition to carrying internal degrees of freedom, e.g. oscillatory localized modes.
When mobilized, these topologically protected excitations permit the transport of mass, energy, charge, spin and other conserved quantities, in a broad range of optical, atomic, soft matter and solid-state systems \cite{BraunKivsharBook,Flach2008Review,PhysRevE.67.056606, PhysRevLett.93.033901,PhysRevLett.97.124101, PhysRevE.73.046602,PhysRevE.76.036603, PhysRevLett.99.214103,PhysRevE.83.036601, PhysRevA.69.053604,malomed2006soliton, PhysRevLett.105.090401,fogarty2015nanofriction,fogarty2016optomechanical,sanchez2012spontaneous,ward2015solid,bohlein2012observation,bylinskii2016observation,kim2012topological,brazovskii2012scanning,roth2015conducting,karpov2016phase}.
Starting with the first theory studies of soliton ratchets, it became clear that the internal modes play a crucial role in the dynamics \cite{quintero2000anomalous,quintero2001anomalies,salerno2002soliton,flach2002broken,morales2003internal,willis2004soliton,martinez2008disorder,cuevas2010regular}. 
In particular the modes in general can couple to external periodic excitations, and due to nonlinearities, induce the motion of the soliton. 
Furthermore, in lattice systems, an effective pinning potential for discrete solitons appears, the so-called Peierls-Nabarro (PN) potential. 
An ab initio theory of Brownian discrete soliton ratchets is nontrivial and currently limited to 1D systems with a continuum limit \cite{sanchez2014collective}.
Experimentally accessible are continuum soliton ratchets in Josephson junction devices \cite{PhysRevE.61.2257,PhysRevLett.87.077002,PhysRevLett.93.087001,PhysRevLett.94.177001,PhysRevLett.95.090603}, while proposals exist for optical \cite{gorbach2006optical}, atomic \cite{PhysRevLett.101.150403} and solid state systems \cite{Zolotaryuk2011}. 

In this Letter, we demonstrate experimental spectroscopy of internal vibrational modes of a micrometer-scale discrete soliton. 
We find that the discrete soliton is capable of rectifying a simple harmonic drive of high frequency, that has a negligible effect in the absence of the soliton. 
Energy from the transverse drive is converted by a nonlinear mechanism to heat, feeding the low frequency localized mode and propelling the quasi-1D soliton along the crystal axis. 
At the presence of damping and fluctuations, we show that the discrete soliton can be directed towards one end of its channel at a rate  conditional on its topological charge and controllable by global external potentials. 
Therefore the presented mechanism could serve as a model for soliton-based transport of mass, electric charge or other conserved quantities.



Trapped ions are well suited for studying fundamental concepts down to the quantum level, featuring unique control in preparation, manipulation, and detection of electronic and motional degrees of freedom \cite{Wineland1979}. 
Isolated in ultrahigh vacuum, they can be laser cooled to micro-Kelvin temperatures and localized to nanometer scale.
The effective potential for an ion near the center of a radiofrequency (RF) Paul trap is approximately harmonic in 3D, with characteristic trapping frequencies $\omega_{\{x,y,z\}}$.
Considering multiple ions, the potential has to be supplemented by the mutual Coulomb interaction  and an ordered, self-assembled crystal is formed that can be scaled to a mesoscopic size of interest to investigate many-body physics \cite{Blumel1988}. 
Figure 1 shows schematically how, for appropriate trapping frequencies $\omega_x\ll\omega_y < \omega_z$, the doubly degenerate ground state of such a crystal takes a planar, inhomogeneous zigzag configuration (we denote the mirrored configuration by \ovl{zigzag}). 
To realize both configurations in one crystal, a localized interface, a domain wall [e.g. the `kink' and `\ovl{kink}' shown in fig.~\ref{fig:kinks}(b)], must form, with a higher energy and the properties of a topological soliton \cite{Landa2010}. Such discrete solitons have been recently characterized theoretically \cite{Campo2010,Chiara2010} manipulated experimentally \cite{Schneider2012a,Mielenz2013,Pyka2013,Ulm2013,Etjemaee13,Landa2013,Partner2013}, and are predicted in circular \cite{Landa2010,MorigiMultipoleRings,Landa2014} and helical configurations \cite{nigmatullin2016formation,zampetaki2015dynamics}. 
In particular, they are proposed to permit quantum coherent manipulation of their internal modes \cite{Marcovitch2008,Landa2013,Landa2014} using the rich toolbox of quantum optics developed for trapped ions \cite{Wineland2012}.

\begin{figure}[h!tb]
	\centering
		\includegraphics[width=3.25in]{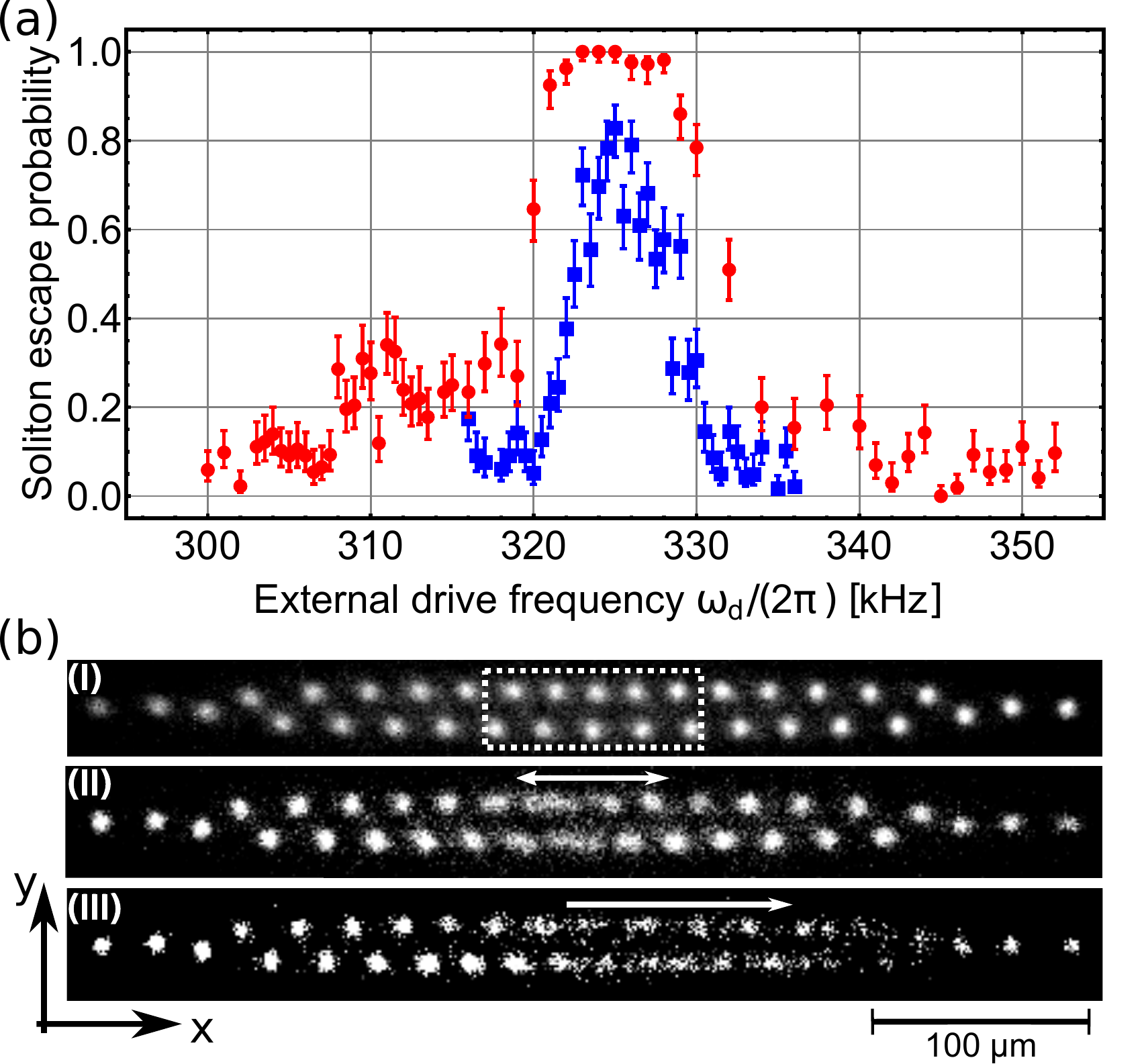}
	\caption{Spectroscopy and directed escape of a discrete soliton.
	\textbf{(a)}
	Spectroscopically resolving kink modes via its complete escape from the crystal. 
	With a periodic excitation of amplitude $\epsilon = 1.45  \cdot 10^{-3}$ (blue squares), we resolve a resonance at $\omega_\text{d}= 2\pi\times\left(325.3 \pm 0.2 \right)\,$kHz with a width of $2\pi\times\left(4.8 \pm 0.5 \right)\,$kHz.
	Strong excitation $\epsilon = 1.74  \cdot 10^{-3}$ (red discs) leads to saturation and an additional weaker resonance at $\omega_\text{d}= 2\pi\times \left(311\pm 0.5 \right)\,$kHz of width $ 2\pi\times\left(4.4 \pm 1.8\right)\,$kHz, in agreement with the numerically derived frequencies of internal modes (see text for details).
	Errorbars represent binomial statistics, and the offset level of 0.1 originates from background gas collisions at $\epsilon = 0$, leading to melting and recrystallization (background lifetime $\approx 3.2\,$s).
	\textbf{(b)} CCD images of the fluorescence of 34 ions confined in a Paul trap.
	(I) The center features the \ovl{kink}.
	(II) The modulation of the radial confinement resonantly excites a localized high-frequency vibrational kink mode, and its nonlinear coupling to other modes results in the axial blurring of the oscillating \ovl{kink}.   
	(III) Escape of the soliton via the right-hand side is witnessed as the left half of the configuration remains unaffected, while on the right, the two opposing ion chains have to `slide' atop each other.
	Since the exposure time of the camera, $t_\text{CCD} = 150\,$ms, is long compared to the duration of propagation, the image depicts the superposition of both configurations.
	}
	\label{fig:spec}
\end{figure}

To describe the dynamics of discrete solitons, we start by considering the $3N$ normal modes of $N$ trapped ions, assuming small oscillations around their equilibrium positions. 
We consider one representative realization with $N=34$ and experimentally determined trap frequencies $\omega_{\{x,y,z\}}=2\pi\times \left( \{38.2, 232.3, 293.0\} \pm 0.1\right)\,$kHz. 
For the zigzag configuration we find mode frequencies $\omega_{\{1,\, \dots, \,102\}}^\text{zigzag}$ in the range $2\pi\times 38.2\,$kHz to $2\pi\times 328\,$kHz, while with a kink, the additional nonlinearity broadens the range of $\omega_{\{1,\, \dots, \,102\}}^\text{kink}$ to $2\pi\times 23.2$\,kHz to $2\pi\times 345\,$kHz \footnote{We have verified that the dynamic nature of the Paul trap \cite{rfions}, leads in our setup to only small frequency shifts}.
The \ovl{zigzag} and \ovl{kink} have identical modes. 
A distinct set of internal modes can be attributed to the kink, with the eigenvectors localized at 8 to 10 ions defining the kink. 

\begin{figure}[h!tb]
	\centering 
		\includegraphics[width=3.25in]{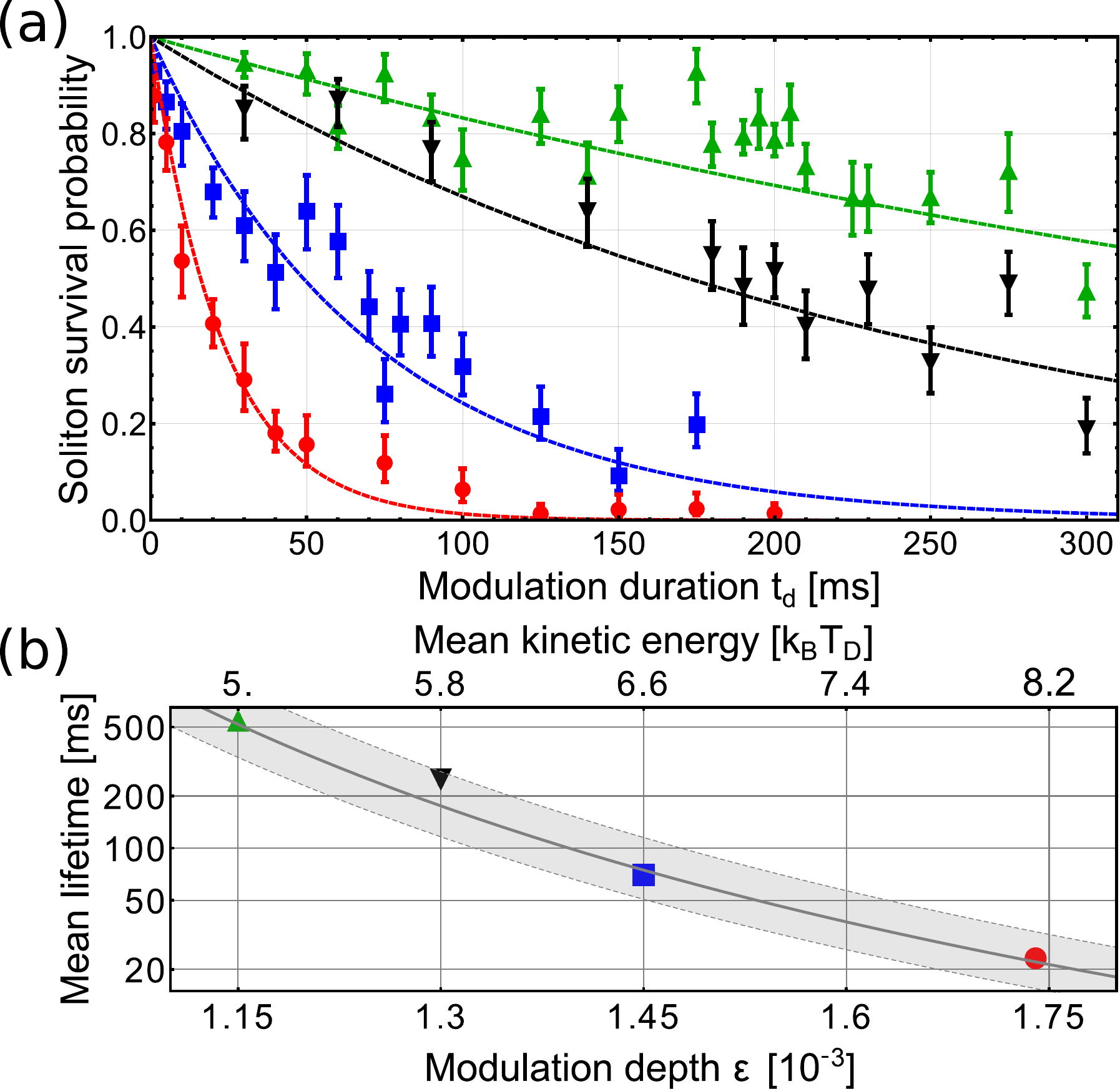}
	\caption{Resonant drive of the radial kink mode leads to a thermally activated escape out of the PN-potential.
	\textbf{(a)} Experimental survival probability of the soliton as function of $t_\text{d}$ for $\omega_\text{d} /(2\pi)= 327\,$kHz  and $\epsilon = 10^{-3}\times\{$1.15\,(green triangles), 1.30\,(black inverted triangles), 1.45\,(blue squares), 1.74\,(red discs)$\}$. 
	Solid lines represent exponential fits yielding the corresponding lifetimes of the kink, $\tau (\epsilon) = \{ (544\pm35)\,\text{ms}, (248\pm16)\,\text{ms}, (71\pm 5)\,\text{ms}, (23\pm2)\,\text{ms}\}$ respectively.
	Errorbars represent the 1$\sigma$ confidence interval. 
	The residual kink loss rate for $\epsilon = 0$ is subtracted based on a calibration measurements.
	\textbf{(b)} Lifetime of the kink in dependence on $\epsilon$ derived from panel (a), fitted with an overdamped Kramers' model \cite{Kramers1940} for symmetric confinement (see text for details), indicated by the gray solid line. 
	The experiment control parameter $\epsilon$ is numerically found to be linearly related to the mean kinetic energy of the ions, leading to an effective temperature. 
	The gray shaded region represents $1\sigma$ uncertainty of the fit.
	We extract a related barrier height of $\left( 26.5 \pm 1.0 \right)\,k_\text{B}T_\text{D}$.
	}
	\label{fig:ex_dur}
\end{figure}

In the experiment, we first probe the internal modes of the discrete soliton, using a spectroscopic  protocol consisting of four steps;
(1) Inducing a phase transition \cite{Blumel1988} from a gas of trapped ions to a Coulomb crystal by laser cooling, (2) in-situ imaging of the crystal to reveal the potential presence of the kink, (3) excitation of the kink using a weak periodic drive, and (4) detecting the configuration, analyzing the kink's response to the excitation.

In step (1), a kink is formed and stabilized \cite{Mielenz2013,Landa2013} with near 0.5 probability.
(2) The crystallized ions scatter laser photons that are collected in a charge-coupled device (CCD) camera, resolving the ion-ion separation with sub-micrometer accuracy, 
allowing to differentiate the crystal configurations.
(3) During an excitation of duration $t_\text{d}$, we modulate the peak voltage ($U_\text{RF}$) on the trap's RF electrodes applying a voltage $U_\text{d}\,\text{sin} \left( \omega_\text{d}\,t \right)$ calibrated by the experimentally determined transfer function of the RF circuit.
Defining the relative excitation depth $\epsilon=U_\text{d}/U_\text{RF}$, the force acting on each ion is derivable from the potential
\begin{equation}
V_\text{d} \propto\epsilon\sin\left( \omega_\text{d}\,t \right) \left[y^2-z^2\right].
\label{eq:Vdrive}\end{equation}
The constant of proportionality is set by the experimental setup, whereas $\epsilon$ remains fully controllable. 
The excitation acts uniformly on the ions' radial coordinates $y$ and $z$, while they remain continuously Doppler cooled by a beam tilted by $<5^\circ$ from the axial direction $x$. 

We run this sequence for crystals of \Mg ions, at least 100 times for each datapoint.
Figure \ref{fig:spec}\,(a) shows two main resonances where the kink [fig.~\ref{fig:spec}(b)(I)] escapes from the crystal, when scanning $\omega_\text{d}$ up to the highest mode frequencies for $t_\text{d} = 85\,$ms. 
These two resonances are close (within their width) to the frequencies of two internal modes of the kink,  $\omega_{97}^\text{kink}$ and $\omega_{100}^\text{kink}$, derived for $\epsilon \rightarrow 0$.
The potential of eq.~\eqref{eq:Vdrive} excites the normal modes, and the reduced peak amplitude for $\omega_{97}^\text{kink}$ results from a smaller projection of its eigenvector components for each ion on the radial direction. 
To shed light on the mechanism leading to the disappearance of the kink at resonance, we image the kink during the driving time $t_\text{d}$.
With small $\epsilon$ and $\omega_\text{d}$ chosen at a resonance, we find an axially blurred trace [fig.~\ref{fig:spec}\,(b)(II)].
We identify this as an induced excitation of the lowest-frequency kink mode, $\omega_1^\text{kink}$, a localized shear mode of the two opposing ion chains oscillating $\pi$-out of phase, capable of axially translating the soliton. 
Increasing $\epsilon$ leads to the dynamics imaged in fig.~\ref{fig:spec}\,(b)(III), demonstrating that due to the radial drive the kink reaches the axial edge of the crystal and escapes, while the crystal as a whole remains intact.

Detailed Molecular Dynamics (MD) simulations confirm these conclusions. 
In addition it reveals that in the limit of vanishing damping and kinetic energy, the individual localized mode resonant with $\omega_\text{d}$ is parametrically excited, and, on a slower timescale comparable with $1/\omega_1^\text{kink} \approx 50\,\mu$s, the energy leaks via nonlinear coupling to the rest of the modes. 
Additionally considering laser cooling along the $x$-axis, the dynamics of a single ion at the low-temperature limit of Doppler cooling can be modeled as a Brownian harmonic oscillator \cite{WinelandReview,marciante2010ion,javanainen1980,javanainen1980a}. 
The damping coefficient $\gamma_x$ and diffusion coefficient $D_x$ are determined by the experiment parameters and obey a fluctuation-dissipation relation $ D_x=\gamma_x k_{\rm B} T_x/m$.
The temperature $T_x$ is of the order of the Doppler-cooling limit $T_{\rm D}\approx 1\,$mK and $\gamma_x/m\approx 2\pi\times 0.3$\,kHz. 
Similar equations hold for the radial coordinates, with $\gamma_y,\gamma_z \ll \gamma_x$.
Adding these Langevin dynamics to the MD simulations including the trap, drive, and Coulomb interactions of all ions, reveals that a nonequilibrium steady state is reached on a millisecond timescale.
The characterization of this state is nontrivial \cite{inprep}, however the mixing of spatial directions by the quasi-2D crystal modes leads to a very effective radial damping, and a mean kinetic energy in the crystal at the steady state, $E_k$, can be defined and is linearly related to $\epsilon$.
Furthermore, using the measured experiment parameters, the shape and position of the resonances in fig.~\ref{fig:spec} are reproduced quantitatively \cite{inprep}.


To further investigate the dynamics of the escape, we experimentally determine the survival probability of the kink as function of $t_\text{d}$, for different values of $\epsilon$, with $\omega_\text{d}$ resonant at $\omega_{100}^\text{kink}$.
The survival probabilities shown in fig.~\ref{fig:ex_dur}(a) can be fitted by an exponential decay yielding a mean lifetime $\tau(\epsilon)$ that decreases with $\epsilon$  [fig.~\ref{fig:ex_dur}(b)] \cite{Partner2015}.
This evidences a thermal activation mechanism for the kink's escape across a barrier. 
As it is known from numerical simulations that the PN potential in a trap becomes effectively harmonic \cite{Mielenz2013}, the value of the PN potential at the edges of the crystal defines the height of the barrier $W$.
To model the dynamics of the escape, we use the numerically obtained positions of all ions to define an instantaneous collective kink coordinate \cite{PhysRevB.38.6713,PhysRevE.48.4037}, adapted from \cite{Partner2013}, for the kink's axial position along the quasi-2D crystal featuring its 1D track. 
Following its time evolution we find that it is overdamped, by using its time derivative to define the velocity and the velocity autocorrelation function. 
We fit the effective damping rate in the experimentally relevant temperature regime \cite{inprep} originating from phonon scattering, finding $g(E_k) \propto E_k^{1/2}$.
Then, assuming that the effective kink coordinate is subject to thermal noise at an effective temperature \footnote{More elaborate models could be considered in a second step \cite{burada2012escape,schuecker2015modulated, geiseler2016kramers}.} defined by $k_B T/2=E_k/(3N)$, we apply Kramers' model in the overdamped limit \cite{Kramers1940,hanggi1990reaction} to describe  the motion of the soliton \cite{grigolini1989brownian,lomdahl1985davydov}.
With a barrier height $W$, the predicted mean lifetime of the Brownian particle is
\begin{equation}
\tau \propto g(T)\,e^{{W}/{(k_\text{B}T)}},\label{eq:Kramers}
\end{equation} 
with the proportionality constant depending on details of the potential well, independent of $T$. 
This approach yields the lifetime averaged over the kink and \ovl{kink} and both directions, and gives $W=\left( 26.5 \pm 1.0\right)\,k_\text{B}T_\text{D}$ based on the experimentally determined $\tau(\epsilon)$ [fig.~\ref{fig:ex_dur}\,(b)].

\begin{figure}[tb]
	\centering
		\includegraphics[width=3.25in]{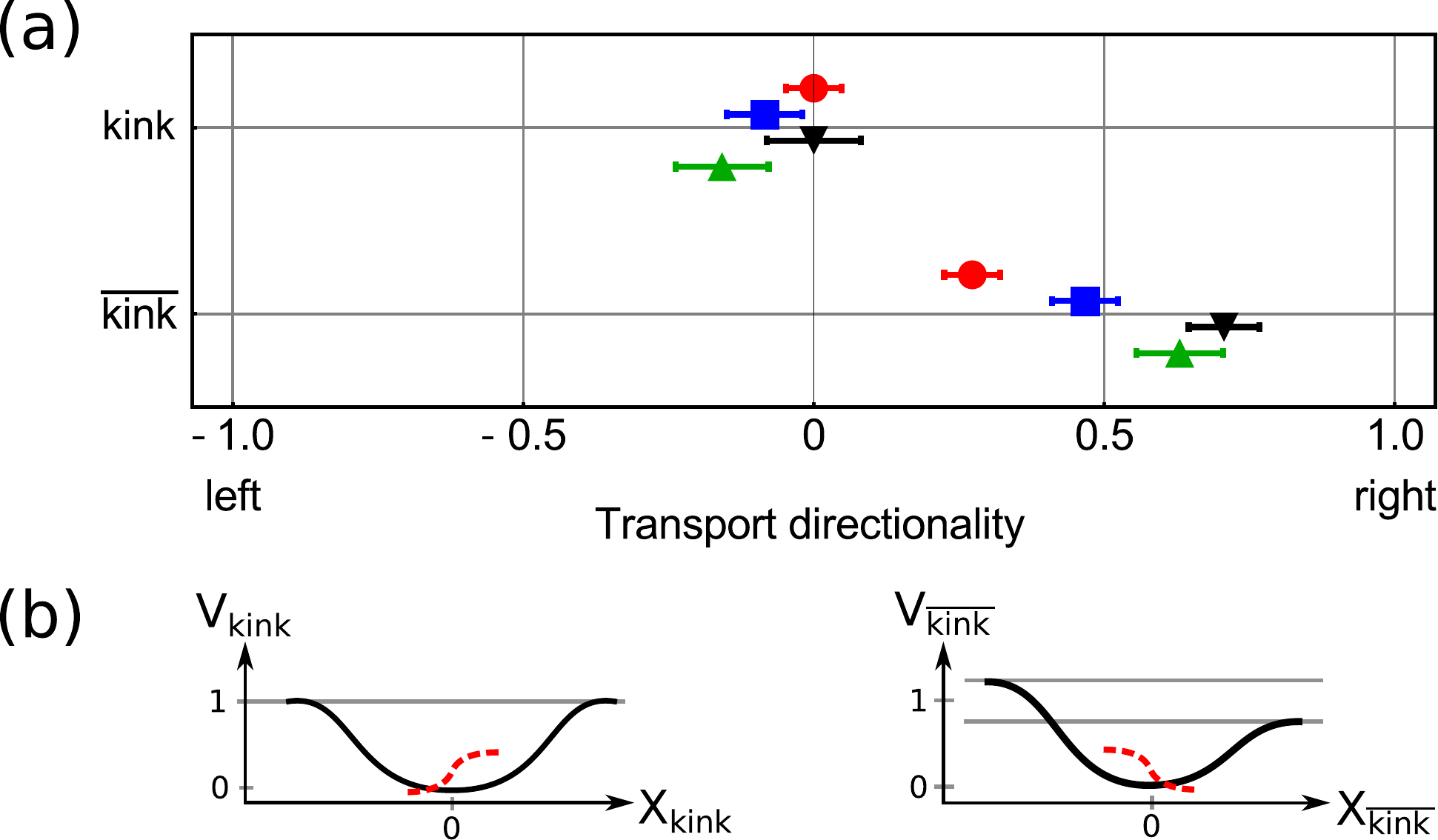}		
	\caption{Experimental transport directionality (TD) conditional on the topological charge: $\pm 1\; \widehat{=}$ \{kink, \ovl{kink}\}.
	\textbf{(a)} While the kink escapes with close to zero TD in the experiment, the \ovl{kink} reveals the broken symmetry tunable by nonlinear terms of the confining potential.
	Additionally, the TD depends on $\epsilon$ (symbols as in fig.~\ref{fig:ex_dur}).
	Errorbars give 1$\,\sigma$ standard deviation.
	\textbf{(b)} The trapping potential for the soliton (schematically depicted) depends on its topological charge. 
	It remains approximately left-right symmetric for the kink and is strongly asymmetric for the \ovl{kink}. 
	The trap depth for the whole Coulomb crystal amounts to $10^4 \times V_\text{kink}$.
	}
\label{fig:direction}
\end{figure}

Finally, we experimentally find a substantial directionality of the soliton transport dependent on its topological charge.
We define the transport directionality (TD) as the difference of probabilities to escape to the right and to the left, normalized by their sum. 
The TD of the kink remains close to zero for all $\epsilon$ [fig.~\ref{fig:direction}\,(a)], while for the \ovl{kink} we find a substantial bias to the right. 
The existence of a mean current requires a broken symmetry.
We extend the description of the harmonic trap potential, accurate at the center of the trap, by nonlinear terms \cite{OzeriDuffing} of third order along the $x$-axis ($L_x$) and $y$-axis ($L_y$) and also fourth order and mixed terms.
The charge density in the trap is sensitive to these terms and we exploit the positions of the ions as a sensor, by minimizing the weighted least-mean-square shift of the imaged ion positions and the measured frequencies $\omega_x$ and $\omega_y$, from their numerically obtained values as a function of the nonlinear coefficients. 
In particular, we find $L_x >0$, which leads to a shift of the whole crystal towards $x<0$, increasing the left-side PN barrier and decreasing it on the right. 
Furthermore, $L_y<0$ shifts the crystal to $y>0$, and due to the different radial densities of the kink and \ovl{kink}, results in different PN barriers.
The mean PN barrier numerically obtained is $W=25.3\,k_\text{B}T_\text{D}$, coinciding within errorbars with the experimental value. 
An intricate interplay of the various global nonlinearity parameters explains the directionality measured in fig.~\ref{fig:direction}, and we obtain an asymmetric shift of about $2\,k_\text{B}T_\text{D}$ for $W$ on the left and on the right. 
This differential shift is comparable to the increase of $T$ with $\epsilon$, which reduces the soliton's sensitivity to the differences in the height of these barriers, as evidenced in fig.~\ref{fig:direction}.
Thus, the directionality can be controlled via the nonlinear terms of the global trapping potential and the amplitude of the drive.

To conclude,
the external radial drive can be tuned to couple selectively to a kink mode, and energy is drawn and converted to heat by the soliton. 
This establishes a power transfer, leading to a nonequilibrium steady state given the laser damping. 
However, despite being microscopically out-of-equilibrium, we find that the axial motion of the soliton can be described by integrating out all degrees of freedom leaving one effective coordinate. 
The directed transport mechanism (that is the manifestation of the underlying nonequilibrium conditions) arises as a consequence of different barrier heights, conditioned on the topological charge, entering Kramers' model, with an effective temperature.
Typically, realizing a ratchet mechanism requires asymmetric gradients at the single particle scale where the combination of nonlinearity, noise, and nonequilibrium drive raise challenges for an efficient control of the transport. 
In this work we show how a large scale potential permits the robust control of the soliton transport, its direction and its rate.

These physical processes connect to a broad range of recent work; among those are investigations of nonequilibrium states \cite{dieterich2015single, grosberg2015nonequilibrium,fodor2016far}, 
the physics of single-particle ratchets in granular chains \cite{berardi2013directed}, in quantum systems \cite{ermann2015quantum,grossert2016experimental}, with power law interactions \cite{liebchen2015interaction}, and
recent experiments and theory studies with trapped ions concerning thermal activation \cite{PhysRevA.83.063401}, escape dynamics \cite{petri2011correlations}, and
steady state heat current formation \cite{lin2011equilibration,bermudez2013controlling, ramm2014energy,freitas2015heat}.
The unique controllability of trapped ions further permits investigations of the dynamics of transport. 
Controlling the global potential allows concatenating crystals along a linear axis with a broken spatial symmetry, or studying ring-configurations \cite{Landa2010,Landa2014}, providing periodic boundary conditions \cite{Schramm2001,Schramm2002}.
Ground state cooling of internal modes enables accessing the quantum regime in mesoscopic, quasi-2D crystals \cite{Landa2010,Landa2014}.
In addition, trapping of several discrete solitons has been realized \cite{Landa2013, Partner2013}. 
Kink mode excitation has been demonstrated via intensity modulation of a laser beam focused on an individual ion of the crystal \cite{inprep}. 
This enables the study of energy transport between kink-lattices, mediated via phonons.

Related work on local kink mode spectroscopy in Coulomb crystals was performed in the context of friction and Aubry transitions \cite{Kiethe2017}.

\begin{acknowledgments}
H.L. thanks Martin Lenz, Ananyo Maitra, and Guglielmo Saggiorato for very fruitful discussions.
H.L. acknowledges support by a Marie Curie Intra European Fellowship within the 7th European Community Framework Programme.

\end{acknowledgments}


\bibliographystyle{unsrt}


\end{document}